\documentstyle[12pt]{article}
\date{}
\begin{document}

\title{\bf Exchange current contributions to the charge radii of the nucleons}

\author{C. Helminen$^*$}
\maketitle

\centerline{\it Department of Physics, University of Helsinki,
00014 Finland}

\vspace{1cm}

\centerline{\bf Abstract}
\vspace{0.5cm}

The exchange charge density operators that correspond to the Fermi-invariant 
decomposition of quark-quark interactions have been constructed. Their 
effect on the electromagnetic charge radii of the nucleons, in combination 
with that of the relativistic corrections to the single-quark operator, 
has been studied with a constituent quark model with a spin and flavor 
dependent hyperfine and a linear confining interaction, which 
gives a quantitative description of the spectra for the light and strange 
baryons. The model gives proton and neutron charge radii in approximate 
agreement with the empirical results assuming reasonable values for the 
radii of the constituent quarks.\\
\vfill
$^*$email address: chelmine@pcu.helsinki.fi
\newpage

\centerline{\bf I. INTRODUCTION}
\vspace{0.5cm}

The charge radii, along with electromagnetic form factors and magnetic 
moments, are observables that provide insight into the internal structure 
of the nucleons. From the measured electric form factors $G_E$ 
of the nucleons the charge radius for the proton is found to be
(in fm) $0.805\pm 0.011$ \cite{Han}, $0.81\pm 0.04$ \cite{Mur}, 
$0.862\pm 0.012$ \cite{Sim1} and $0.88\pm 0.03$ \cite{Bor}. The study of the 
charge form factor of the neutron is experimentally difficult due to the 
absence of pure neutron targets. Electron-deuteron scattering and scattering 
of slow neutrons off atomic electrons have, however, shown the mean square 
charge radius of the neutron to be nonzero and negative. Typical
experimental values for the 
mean square neutron charge radius are (in fm$^2$), e.g.,  
$-0.113\pm 0.003\pm 0.004$ \cite{Kop} and $-0.117\pm 0.002$ \cite{Dzi} 
(when averaging \cite{Kro} and \cite{Koe}). Since the electric Sachs
form factor $G_E$ is defined as a linear combination of the Dirac and
Pauli form factors $F_1$ and $F_2$ the charge radius of the
nucleon can be divided into an "intrinsic" part, coming from $F_1$, and 
an "anomalous magnetic moment" part, coming from $F_2$ \cite{Fol}.
In the case of the neutron the main part of the charge radius in fact
comes from this "Pauli" term, while the "Dirac" term will be small and 
positive \cite{EW}. Quark model calculations of the charge radii of
the nucleons concern the intrinsic part. Due to the small mass of
the (constituent) quarks relativistic corrections to the charge density 
operator should also be taken into account in the impulse approximation.
\par

In interquark interaction models that are flavor or velocity dependent 
(the interaction suggested in Ref. \cite{GR} being an example of the former)
exchange current contributions will arise as a consequence of the continuity 
equation. The continuity equation links the terms in the the charge
density operator of any given order in $(v/c)$ to terms of the next
order in $(v/c)$ in the interaction. The most obvious constraint is that
two-body contributions in the charge density have to have a vanishing
volume integral.
Exchange current contributions to the current operator have been included,
e.g., in the work related in Ref. \cite{BHY1}, where pion and gluon 
exchange current contributions were calculated and in Ref. \cite{BHY2}, 
where also exchange current contributions
from the confinement current were taken into account.
\par

It is possible to express the quark-quark interaction  
in terms of the five relativistic Fermi spin invariants $SVTAP$ \cite{Gold}, 
corresponding to effective scalar, vector, tensor, axial vector and 
pseudoscalar exchange. The corresponding exchange charge density operators
associated with these invariants have been derived in the present work
for light and strange (SU(3)) quarks. The exchange current contributions 
to the charge radius of the nucleons have then been calculated using a 
recently developed phenomenological model for the quark-quark interaction, 
which in combination with a static linear confining interaction gives a good 
description of the spectra of the light and strange baryons \cite{GPVW}.
\par

The possibility that the constituent quarks would have a non-trivial
electromagnetic structure, described by constituent quark form factors, has 
also been explored in this work. The total nucleon charge form
factor, and subsequently the charge radius, would then also get 
contributions from the quark form factors.
It is found that the quark model interaction of Ref. \cite{GPVW} leads
to exchange current contributions, which, when combined with the impulse
approximation and the anomalous magnetic moment part of the nucleon
charge radii, while assuming reasonable values for the radii of the
constituent quarks, give satisfactory values for the 
charge radii of the nucleons.
\par

This paper is divided into six sections. Section II defines the charge
form factor and the charge radius of the nucleon,
taking into account relativistic corrections. In section III
the charge form
factor and the charge radius of the nucleon 
is derived with inclusion of exchange current
contributions, and in section IV these results are applied to the 
phenomenological model of Ref. \cite{GPVW}. The possible charge form
factor and radius of the constituent quarks, and their effect on the
charge radii of the nucleons are discussed in section V, and a summarizing
discussion is given in section VI. 
\vspace{1cm}

\centerline{\bf II. THE CHARGE FORM FACTOR}
\centerline{\bf AND THE CHARGE RADIUS}
\vspace{0.5cm}

The Sachs charge form factor $G_E(q^2)$ of a nucleon is defined as

$$G_E(q^2)=F_1(q^2)-{q^2\over 4m_N^2}F_2(q^2)\ ,\eqno(2.1) $$

\noindent where $F_1(q^2)$ is the Dirac form factor and $F_2(q^2)$ the
Pauli form factor. The form factors are normalized as $F_1(0)=Q$, i.e., the
electric charge of the nucleon in units of $e$, and $F_2(0)=\kappa$, 
i.e., the anomalous magnetic moment of the nucleon in units of
$({e\over2m_N})$. The (mean square) charge radius of the nucleon can be
calculated from the charge form factor as 

$$<r^2>_N=-6\ {dG_E(q^2)\over d(q^2)}\Big|_{q^2=0}
=-6\Big\{
{dF_1(q^2)\over d(q^2)}\Big|_{q^2=0} - {F_2(0)\over 4m_N^2}\Big\}$$
$$= -6{dF_1(q^2)\over d(q^2)}\Big|_{q^2=0}+{3\over2}{F_2(0)\over m_N^2}
\ ,\eqno(2.2)$$

\noindent The mean square charge radius is thus divided into two
parts, $<~r^2~>_N$ = $<~r^2~>_{int,N}+<~r^2~>_{an,N}$, where the
first term comes from the intrinsic charge distribution of the
nucleon while the second term arises from the anomalous magnetic
moment of the nucleon, which for the proton is $\kappa_p=1.793$ and
for the neutron is $\kappa_n=-1.913$. Using
$F_{2,p}(0)=\kappa_p$, $F_{2,n}(0)=\kappa_n$ and $m_N=939$ MeV,
the results are $<r^2>_{an,p}=0.119$ fm$^2$ and 
$<r^2>_{an,n}=-0.127$ fm$^2$. The remaining (intrinsic) part of $<r^2>_N$
is interpreted as coming from
the charge density of a system of three constituent quarks.
\par

If a constituent quark (denoted here as $i$) is treated as a point Dirac 
particle without anomalous terms, its electromagnetic current operator can 
be expressed as

$$<p'_i|J_\mu(0)|p_i>=i\bar u(p_i')\gamma_\mu Q^{(i)}u(p_i)\ ,\eqno(2.3)$$

\noindent when $p_i$ and $p_i'$ denote the initial and the final momenta, 
respectively, and 

$$Q^{(i)}={1\over 2}[\lambda_3^{(i)}
+{1\over\sqrt{3}}\lambda_8^{(i)}] \eqno(2.4)$$ 

\noindent (in units of charge $e$) is the (SU(3)) 
charge operator of the quark $i$. An argument for the 
absence of anomalous current terms is given in \cite{Di}.
Since $J_\mu=({\bf J},i\rho)$ the electromagnetic charge density  operator is 

$$<p'_i|\ \rho\ |p_i>=\bar u(p'_i)\gamma_4Q^{(i)}u(p_i)$$
$$\qquad\qquad=\sqrt{{(E'_i+m)(E_i+m)\over 4E_i'E_i}}[1+
{{{\bf p}_i}'\cdot{\bf p}_i+i\mbox{\boldmath$\sigma$}^{(i)}\cdot{{\bf p}_i}'
\times{\bf p}_i\over (E_i'+m)(E_i+m)}]Q^{(i)}\ ,\eqno(2.5)$$

\noindent where $m$ is the mass of the constituent quark and 
$E_i=\sqrt{{{\bf p}_i}^2+m^2}$. By introducing a velocity operator ${\bf v}_i=
{1\over 2m}({{\bf p}_i}'+{\bf p}_i)$ and the momentum transfer ${\bf q}=
{{\bf p}_i}'-{\bf p}_i$, the momentum operators ${\bf p}_i$ and
${{\bf p}_i}'$ can be expressed as

$${\bf p}_i=m{\bf v}_i-{{\bf q}\over 2}$$
$${{\bf p}_i}'=m{\bf v}_i+{{\bf q}\over 2}\ ,\eqno(2.6)$$

\noindent and the expression in Eq. (2.5) for the charge density 
operator will, to lowest order in $q^2$, be 

$$<p_i'|\ \rho\ |p_i>=(1-{q^2\over 8 m^2})\ Q^{(i)}\ . \eqno(2.7)$$

\noindent The above expression in Eq. (2.7) is then the 
impulse approximation with
the (relativistic) Darwin-Foldy correction. The spin-orbit term in Eq. (2.5) 
is linear in ${\bf q}$  and gives no contribution for the
ground state baryons, and has therefore been left out in Eq. (2.7).
A charge form factor (which is interpreted as the Dirac part of the
electric Sachs form factor) can now be calculated as the Fourier 
transform of the charge density operator. 
\par

When using a three-body wave function that is symmetric with respect 
to the combined spin, flavor and spatial coordinates, 
the charge form factor of a system of three 
(constituent) quarks with the same mass $m$ will then in the impulse 
approximation be

$$F_{C,IA}(q^2)=3<Q^{(1)}>\int d^3r_1d^3r_2d^3r_3|\psi({\bf r}_1,{\bf r}_2,
{\bf r}_3)|^2 e^{i{\bf q}\cdot{\bf r}_1}[1-{q^2\over 8 m^2}]$$
$$\qquad=3<Q^{(1)}>\int d^3r_1d^3r_2d^3r_3|\psi({\bf r}_1,{\bf r}_2,
{\bf r}_3)|^2 [1-{q^2r_1^2\over 6}+{\cal O}(q^4)][1-{q^2\over 8 m^2}]\ ,
\eqno(2.8)$$

\noindent where in the expansion of the factor $e^{i{\bf q}\cdot{\bf r}_1}$
only the spatial scalar part needs to be taken into account for a ground state 
baryon. The terms indicated by ${\cal O}(q^4)$ are of higher order in $q^2$.
The impulse approximation (with relativistic corrections) for the 
corresponding mean square charge radius can then be calculated as
$<r^2>_{IA}=-6{dF_{C,IA}\over d(q^2)}|_{q^2=0}$, 
giving

$$<r^2>_{IA}=3<Q^{(1)}>\int d^3r_1d^3r_2d^3r_3|\psi({\bf r}_1,{\bf r}_2,
{\bf r}_3)|^2 [r_1^2+{3\over 4m^2}]\ . \eqno(2.9)$$

\vspace{1cm}

\centerline{\bf III. EXCHANGE CURRENT CONTRIBUTIONS}
\vspace{0.5cm}

The interaction between quarks may be decomposed in terms of Fermi
invariants as

$$V=\sum_{j=1}^5 v_j F_j\ ,\eqno(3.1)$$

\noindent where $F_j=\ S,\ V,\ T,\ A,\ P$, defined as \cite{Gold}

$$ S=1^{(1)}1^{(2)}\ ,\quad V=\gamma_\mu^{(1)}\gamma_\mu^{(2)}\ ,\quad T=
{1\over2}\sigma_{\mu\nu}^{(1)}\sigma_{\mu\nu}^{(2)}\ ,$$

$$A=i\gamma_5^{(1)}\gamma_\mu^{(1)}i\gamma_5^{(2)}\gamma_\mu^{(2)}\ ,\quad
P=\gamma_5^{(1)}\gamma_5^{(2)}\ .\eqno(3.2)$$

\noindent Exchange current corrections to the charge density operator 
$\rho$ for a two-quark system will arise as contact current terms in
the nonrelativistic reduction of the 5 Fermi spin invariants \cite{TRB}.
The exchange charge density operators $\rho_j$, $j=1\dots 5$,
corresponding to the $SVTAP$ decomposition, consist of flavor
independent and flavor dependent parts $\rho_j^+$ and $\rho_j^-$, 
respectively. 
\par

The simplest method for deriving the flavor independent exchange charge 
density term $\rho_1^+$, associated with effective scalar exchange, is 
the following. If $v_1^+(\bf k)$ is the corresponding flavor independent
potential, performing a mass shift $m\to m^*=m+v_1^+$ in the 
relativistic Darwin-Foldy correction term in Eq. (2.7) leads, to 
lowest order in $v_1^+$, to the exchange charge contribution 

$$\rho_1^+({\bf k}_1,{\bf k}_2,{\bf q})=\rho_1^+({\bf k}_2,{\bf q})+
\rho_1^+({\bf k}_1,{\bf q})$$
$$\qquad\qquad={q^2\over 8 m^2}\cdot{2v_1^+({\bf k}_2)\over m}Q^{(1)}+
{q^2\over 8 m^2}\cdot{2v_1^+({\bf k}_1)\over m}Q^{(2)}\ ,\eqno(3.3)$$

\noindent where ${\bf k}_2$ and ${\bf k}_1$ are the momentum transfers
from quark 2 to quark 1 and from quark 1 to quark 2, respectively, while 
${\bf q}$ is the momentum transfer to the two-quark system.
\par

The more general way of constructing all of the exchange charge density
operators $\rho_j$ to lowest order in $(v/c)$ is by
decomposing the quark propagators in the (relativistic) Born
terms in the $\gamma qq\rightarrow qq$ amplitudes into positive and negative
energy components, retaining only the negative ones (cf. 
calculations of exchange current corrections to two-nucleon systems in 
Refs. \cite{TRB} and \cite{BR}), giving

$$\rho_j({\bf k}_1,{\bf k}_2,{\bf q})=\rho_j({\bf k}_2,{\bf q})+
\rho_j({\bf k}_1,{\bf q})$$
$$\qquad\qquad={1\over 8 m^3}O_j(2Q^{(1)}v_j^+({\bf k}_2)+
\lbrace Q^{(1)},\sum_{k=1}^8\lambda_k^{(1)}\lambda_k^{(2)}\rbrace v_j^-
({\bf k}_2))$$
$$\qquad + (1\leftrightarrow 2)\ , \eqno(3.4)$$

\noindent where $v_j^+$ and $v_j^-$ are flavor independent and 
flavor dependent potentials, respectively, and $(1\leftrightarrow2)$ 
is a term with the coordinates of quarks 1 and 2 exchanged.
\par

The operators $O_j$, $j=1\dots5$, corresponding to effective scalar, vector,
tensor, axial vector and pseudoscalar exchange mechanisms, will be

$$O_1=q^2+2i\mbox{\boldmath$\sigma$}^{(1)}\cdot{\bf P}_1\times{\bf q}\ ,
\eqno(3.5a)$$

$$O_2={\bf q}\cdot{\bf k}_2+(\mbox{\boldmath$\sigma$}^{(2)}\times{\bf k}_2)
\cdot(\mbox{\boldmath$\sigma$}^{(1)}\times{\bf q})-
2i\mbox{\boldmath$\sigma$}^{(1)}\cdot{\bf q}\times{\bf P}_2\ ,\eqno(3.5b)$$

$$O_3={\bf q}\cdot{\bf k}_2+\mbox{\boldmath$\sigma$}^{(1)}\cdot
\mbox{\boldmath$\sigma$}^{(2)}{\bf q}\cdot{\bf k}_2-
\mbox{\boldmath$\sigma$}^{(2)}\cdot{\bf q}\ \mbox{\boldmath$\sigma$}^{(1)}
\cdot({\bf k}_2-{\bf q})+2i\mbox{\boldmath$\sigma$}^{(2)}\cdot{\bf P}_2\times
{\bf q}\ ,\eqno(3.5c)$$

$$O_4=q^2\mbox{\boldmath$\sigma$}^{(1)}\cdot\mbox{\boldmath$\sigma$}^{(2)}+
2i\mbox{\boldmath$\sigma$}^{(2)}\cdot{\bf P}_1\times{\bf q}-
\mbox{\boldmath$\sigma$}^{(2)}\cdot{\bf q}\ \mbox{\boldmath$\sigma$}^{(1)}
\cdot{\bf q}+\mbox{\boldmath$\sigma$}^{(1)}\cdot{\bf q}\ 
\mbox{\boldmath$\sigma$}^{(2)}\cdot{\bf k}_2\,
\eqno(3.5d)$$

$$O_5=\mbox{\boldmath$\sigma$}^{(2)}\cdot{\bf k}_2\ \mbox{\boldmath$\sigma$}
^{(1)}\cdot{\bf q}\ .
\eqno(3.5e)$$

\noindent In the expressions above ${\bf P}_1={1\over 2}({\bf p}_1
+{{\bf p}_1}')$ and ${\bf P}_2={1\over 2}({\bf p}_2+{{\bf p}_2}')$, 
where ${\bf p}_1$ and ${\bf p}_2$ are the initial quark momenta, and 
${{\bf p}_1}'$ and ${{\bf p}_2}'$ are the final quark momenta.
\par

For simplicity the following notations are used: 
$\lbrace Q^{(1)},\sum_{k=1}^8
\lambda_k^{(1)}\lambda_k^{(2)}\rbrace\equiv Q^{(12)}$,
$\lbrace Q^{(2)},\sum_{k=1}^8\lambda_k^{(2)}\lambda_k^{(1)}\rbrace$
$\equiv Q^{(21)}$. The operator $Q^{(ij)}$ can be cast in the form

$$Q^{(ij)}={2\over 3}(\lambda_3^{(j)}+{1\over\sqrt{3}}\lambda_8^{(j)})+
{1\over\sqrt{3}}(\lambda_8^{(i)}\lambda_3^{(j)}+
\lambda_3^{(i)}\lambda_8^{(j)})$$
$$\qquad\qquad+{1\over 3}
\sum_{k=1}^5\lambda_k^{(i)}\lambda_k^{(j)}-{2\over 3}\sum_{k=6}^7
\lambda_k^{(i)}\lambda_k^{(j)}-{1\over 3}\lambda_8^{(i)}\lambda_8^{(j)}\ .
\eqno(3.6)$$

The general form for an exchange current contribution to the charge 
form factor can be expressed as

$$F_{C,ex}(q^2)=3<\int {d^3k_1\over(2\pi)^3}{d^3k_2\over(2\pi)^3}
e^{i({\bf k}_1\cdot{\bf r}_1+{\bf k}_2\cdot{\bf r}_2)}(2\pi)^3\delta^{(3)}
({\bf k}_1+{\bf k}_2-{\bf q})\rho_{ex}({\bf k}_1,{\bf k}_2,{\bf q})>$$
$$\qquad\qquad\qquad=3<e^{i{\bf q}\cdot{\bf r}_1}\int {d^3k_2\over(2\pi)^3}
e^{-i{\bf k}_2\cdot{\bf r}_{12}}\rho_{ex}({\bf k}_2,{\bf q})>
+(1\leftrightarrow 2)\ ,\eqno(3.7)$$

\noindent where $\rho_{ex}$ is the exchange charge density operator and 
${\bf r}_{12}={\bf r}_1-{\bf r}_2$. The corresponding contribution to
the mean square charge radius is then 
$<r^2>_{ex}=-6{dF_{C,ex}\over d(q^2)}|_{q^2=0}$.
The total (intrinsic) charge form factor is then $F_C=F_{C,IA}\cdot F_{C,ex}$, 
i.e., 

$$(1-{r^2q^2\over 6}+{\cal O}(q^4) )=(1-{r^2_{IA}q^2\over 6}+{\cal O}(q^4) )
(1-{r^2_{ex}q^2\over 6}+{\cal O}(q^4) )$$
$$=(1-{(r^2_{IA}+r^2_{ex})q^2\over 6}+{\cal O}(q^4) )\ ,\eqno(3.8)$$
 
\noindent giving an intrinsic mean square charge radius defined as 
$<r^2>_{int}=<r^2>_{IA}+<r^2>_{ex}$.
\par

Taking into account only the spatial
scalar component of the exchange charge density operators for a ground state
baryon, and noting that the matrix elements of terms in Eq. (3.5) that contain 
${\bf P}_1$ or ${\bf P}_2$ will be small, the expressions for the charge
density operators $\rho_j$ can be simplified. The contribution to the 
charge density operator from scalar meson exchange mechanisms will then be

$$\rho_1({\bf k}_1,{\bf k}_2,{\bf q})=
{1\over 8m^3}q^2\ [2Q^{(1)}v_1^+({\bf k}_2)+Q^{(12)}v_1^-({\bf k}_2)]
+(1\leftrightarrow 2)\ .\eqno(3.9)$$

\noindent For vector meson exchange mechanisms the corresponding
expression is

$$\rho_2({\bf k}_1,{\bf k}_2,{\bf q})=
{1\over 8m^3}[{\bf q}\cdot{\bf k}_2+
2{\bf q}\cdot{\bf k}_2{\mbox{\boldmath$\sigma$}^{(1)}
\cdot\mbox{\boldmath$\sigma$}^{(2)}\over 3}]
[2Q^{(1)}v_2^+({\bf k}_2)+Q^{(12)}v_2^-({\bf k}_2)]$$
$$\qquad\qquad\qquad + (1\leftrightarrow 2)\ ,\eqno(3.10)$$

\noindent and the exchange charge 
density operator for effective tensor exchange will be

$$\rho_3({\bf k}_1,{\bf k}_2,{\bf q})={1\over 8m^3}[{\bf q}\cdot{\bf k}_2+
(2{\bf q}\cdot{\bf k}_2+q^2){\mbox{\boldmath$\sigma$}^{(1)}
\cdot\mbox{\boldmath$\sigma$}^{(2)}\over 3}]$$
$$\qquad [2Q^{(1)}v_3^+({\bf k}_2)+Q^{(12)}v_3^-({\bf k}_2)]
 + (1\leftrightarrow 2)\ .\eqno(3.11)$$

\noindent The contribution to the exchange charge density operator from 
axial vector exchange mechanisms is derived as

$$\rho_4({\bf k}_1,{\bf k}_2,{\bf q})={1\over 8m^3}[2q^2+
{\bf q}\cdot{\bf k}_2]{\mbox{\boldmath$\sigma$}^{(1)}\cdot
\mbox{\boldmath$\sigma$}^{(2)}\over 3}
[2Q^{(1)}v_4^+({\bf k}_2)+Q^{(12)}v_4^-({\bf k}_2)]$$
$$\qquad\qquad\qquad + (1\leftrightarrow 2)\ ,\eqno(3.12)$$

\noindent and, finally, the contribution from 
effective pseudoscalar meson exchange  can be derived as

$$\rho_5({\bf k}_1,{\bf k}_2,{\bf q})={1\over 8m^3}{\bf q}\cdot{\bf k}_2
{\mbox{\boldmath$\sigma$}^{(1)}\cdot\mbox{\boldmath$\sigma$}^{(2)}\over 3}
[2Q^{(1)}v_5^+({\bf k}_2)+Q^{(12)}v_5^-({\bf k}_2)]$$
$$\qquad\qquad\qquad + (1\leftrightarrow 2)\ .\eqno(3.13)$$

\vspace{1cm}

\centerline{\bf IV. NUCLEON CHARGE RADII IN THE CHIRAL}
\centerline{\bf CONSTITUENT QUARK MODEL}
\vspace{0.5cm}

As an illustration of how exchange currents may contribute to the total
charge radius of the nucleon the above formalism has been applied to a
quark-quark interaction model that is flavor dependent.
In Ref. \cite{GPVW} a  phenomenological model for the 
hyperfine interactions between quarks, combined with a static linear 
confining interaction, was used to get a satisfactory description of the 
spectra of the light and strange baryons. The Hamiltonian of the model
is of the form

$$H=\sum^3_{i=1}\sqrt{{\bf p}_i^2+m_i^2}+\sum_{i<j}^3V({\bf r}_{ij})\ .
\eqno(4.1)$$

The quark-quark interaction $V({\bf r}_{ij})$ consists of a 
(linear) confining part $V_{conf}({\bf r}_{ij})$, and a hyperfine 
interaction term $V^{octet}_\chi({\bf r}_{ij})$ of the form

$$V^{octet}_\chi({\bf r}_{12})=\left\{\sum_{k=1}^3 V_\pi({\bf r}_{12})
\lambda^{(1)}_k\lambda^{(2)}_k+
\sum_{k=4}^7 V_K({\bf r}_{12})\lambda^{(1)}_k\lambda^{(2)}_k\right.$$

$$\qquad\quad\left. +V_\eta({\bf r}_{12})\lambda^{(1)}_8\lambda^{(2)}_8
\right\}\mbox{\boldmath$\sigma$}^{(1)}\cdot\mbox{\boldmath$\sigma$}^{(2)}
\ ,\eqno(4.2)$$

\noindent where $\lambda_k$ and {\boldmath$\sigma$} are the flavor
and spin matrices of the quarks. In the full interquark potential of Ref.
\cite{GPVW} also a singlet exchange term was included.
This smaller term will not be considered here.
The spatial part of $V^{octet}_\chi$ for the model is taken to 
have the form 

$$V_\gamma({\bf r}_{ij})={g^2_\gamma\over 4\pi}{1\over 3}
{1\over 4m_i m_j}\lbrace\mu^2_\gamma{e^{-\mu_\gamma r_{ij}}\over r_{ij}}-
\lambda^2_\gamma{e^{-\lambda_\gamma r_{ij}}\over r_{ij}}\rbrace\ ,
\eqno(4.3)$$

\noindent \noindent where $\gamma=\pi,K,\eta$, ${g^2_\gamma\over 4\pi}=0.67$
for $\pi,K,\eta$, $m_i$ is the constituent quark mass,
corresponding to either $m_u=m_d=$ 340~MeV or
$m_s=$ 500~MeV, and $\lambda_\gamma=\lambda_0+
\kappa\mu_\gamma$, with $\lambda_0=2.87$~fm$^{-1}$ and $\kappa=0.81$.
\par

This model (one version of the so called chiral constituent 
quark model) will be used below, combined with an approximation of the
three-body wave function of Ref. \cite{GPVW}. The 
wave function approximation is of the form 
$\psi({\bf r}_1,{\bf r}_2,{\bf r}_3)=({m\omega\over\pi})^{3/2}
e^{i{\bf P}\cdot{\bf R}}e^{-{m\omega\over 2}(r^2+\rho^2)}$, with an 
effective oscillator frequency $\omega=$1240~MeV, 
corresponding to an estimate of the impulse 
approximation radius without relativistic corrections in the model of 
Ref. \cite{GPVW}. The coordinates ${\bf R}$, ${\bf r}$ and 
{\boldmath$\rho$} are defined as 

$${\bf R}={{\bf r}_1+{\bf r}_2+{\bf r}_3\over\sqrt{3}}$$

$${\bf r}={{\bf r}_1-{\bf r}_2\over\sqrt{2}}={{\bf r}_{12}\over\sqrt{2}}$$

$$\mbox{\boldmath$\rho$}={{\bf r}_1+{\bf r}_2-2{\bf r}_3\over\sqrt{6}}
\ .\eqno(4.4)$$
 
\noindent  In the calculations the (renormalized) center-of-mass coordinate 
${\bf R}$ is removed. With this wave function the impulse approximation 
(with relativistic corrections) for the (intrinsic) mean square charge 
radius can be calculated from Eq. (2.9) as 

$$<r^2>_{IA}=3<Q^{(1)}>[{1\over m\omega}+{3\over 4m^2}]\ .\eqno(4.5)$$

The static linear confining interactions of Ref. \cite{GPVW}
is spin and flavor independent, and can formally be viewed as a static 
approximation to a scalar exchange interaction, the sign of which is 
positive instead of negative as for a conventional scalar exchange interaction 
\cite{BHY2}. In the model of Ref. \cite{GPVW} the confining potential is

$$\tilde v_{conf}({\bf r}_{12})=Cr_{12}+V_0\ ,\eqno(4.6)$$

\noindent where the value of the parameter $C$ is 2.33~fm$^{-2}$ and where
$V_0=-416$~MeV. With the approximated wave function the confinement 
contribution to the charge form factor can be calculated from Eq. (3.7) and
Eq. (3.9), where $v_1^+({\bf k})$ is the inverse Fourier transform of the 
confining potential. The corresponding mean square charge radius will then be

$$<r^2>_{conf}=-{9\over m^3}[\sqrt{2\over\pi m\omega}C+{V_0\over2}]
<Q^{(1)}+Q^{(2)}>R(\omega)\ .\eqno(4.7)$$

\noindent The factor $R(\omega)$ represents a relativistic correction to the
exchange charge density operator, originating in the spinor normalization
factor and the energy denominator in the small component of the quark wave
function (see Ref. \cite{DGHR} for a discussion on this factor in the 
context of exchange magnetic moment operators), the value of which is 0.28
for $\omega=1240$~MeV. The confinement contribution can, as is readily seen
from the above equations, be viewed as a one-body contribution to the charge
radius because of the equality $<Q^{(1)}>=<Q^{(2)}>$.
\par

The potential function 
$v({\bf k})$ of the hyperfine interaction can be obtained as the inverse
Fourier transform of \cite{DGHR}

$$\tilde v_\gamma({\bf r}_{12})={4m^2\over r_{12}}\int_{r_{12}}^\infty dr'
\int_{r'}^\infty dr''r''f_\gamma(r'')\ ,\eqno(4.8)$$

\noindent where $f_\gamma(r)$ is obtained from the relation $V_\gamma
({\bf r}_{12})={1\over 3}f_\gamma(r_{12})$ in combination with 
Eq. (4.3), thus giving

$$f_\gamma(r)={g^2_\gamma\over 4\pi}{1\over 4m^2}\lbrace\mu_\gamma^2
{e^{-\mu_\gamma r}\over r}-
\lambda_\gamma^2{e^{-\lambda_\gamma r}\over r}\rbrace\ .\eqno(4.9)$$

The hyperfine interaction can in this model be 
interpreted as coming solely from pseudoscalar exchange mechanisms, since
the volume integral of the interaction vanishes. 
\noindent The relativistic correction $R(\omega)$ should also be 
taken into account and, using Eq. (3.13), one finally gets for the 
effective pseudoscalar ($P$) meson exchange 
contribution to the mean square charge radius 

$$<r^2>_P={\pi\over
2m^3}({m\omega\over\pi})^{3/2}<\mbox{\boldmath$\sigma$}^{(1)}
\cdot\mbox{\boldmath$\sigma$}^{(2)}\ Q^{(12)}+
\mbox{\boldmath$\sigma$}^{(2)}\cdot\mbox{\boldmath$\sigma$}^{(1)}\ Q^{(21)}>$$

$$\qquad\int_0^\infty dr r^3 e^{-m\omega r^2}{\partial\over
\partial r}\tilde v(\sqrt{2}r)R(\omega)\ .\eqno(4.10)$$

\noindent In the calculations the spin-flavor part of Eq. (4.10)
is divided into terms that correspond to the different potential functions 
$\tilde v_\gamma$ defined in Eq. (4.8).  
This is done by noting that for nucleons the flavor operator $Q^{(ij)}$ in
Eq. (3.6) can be written as $Q^{(ij)}=Q_\pi^{(ij)}+Q_\eta^{(ij)}+
Q_{\pi+\eta}^{(ij)}$, where $Q_\pi^{(ij)}$, $Q_\eta^{(ij)}$ and 
$Q_{\pi+\eta}^{(ij)}$ are associated with the potential functions 
$\tilde v_\pi$, $\tilde v_\eta$ and ${1\over 2}(\tilde v_\pi+\tilde v_\eta)$,
respectively. The matrix elements of the spin-flavor parts of the exchange
current contributions are given in Table I. The numerical values for the
mean square charge radius contributions from the impulse approximation 
(with relativistic correction), from confinement and from pseudoscalar 
exchange mechanisms are given for the proton and the neutron in Table II. \\

\vspace{1cm}

\centerline{\bf V. CHARGE FORM FACTORS OF THE QUARKS}
\vspace{0.5cm}

In the previous sections the constituent quarks have been treated as pointlike 
objects without internal structure. If, on the other hand, the constituent 
quarks have a non-trivial electromagnetic structure, i.e., they are assumed to
be dressed by their mesonic (quark-antiquark) polarization clouds, quark form
factors can be introduced. The charge operator $Q^{(i)}$ in Eq. (2.4) can be 
expressed as 
$Q^{(i)}(0)={1\over 2}F_3(0)\lambda_3^{(i)}+
{1\over 2\sqrt{3}}F_8(0)\lambda_8^{(i)}$,
\noindent where $F_3$ and $F_8$ are possible quark form factors, normalized as 
$F_3(0)=F_8(0)=1$, and the expression for $Q^{(i)}(q^2)$ is then

$$Q^{(i)}(q^2)={1\over 2}F_3(q^2)\lambda_3^{(i)}+{1\over 2\sqrt{3}}F_8(q^2)
\lambda_8^{(i)}\ .\eqno(5.1)$$

\noindent A possible quark contribution to the form factors of nucleons can
subsequently be expressed as 

$$F_{q,p}=3<p|Q^{(1)}|p>={1\over 2}(F_3+F_8)\ ,$$
$$F_{q,n}=3<n|Q^{(1)}|n>={1\over 2}(-F_3+F_8)\ .
\eqno(5.2)$$

\noindent It is possible to re-express 
the quark form factors in terms of contributions from up and down quarks. 
One notes that

$${2\over 3}F_u=<u|Q^{(1)}|u>={1\over 2}F_3+{1\over 6}F_8\ ,$$
$$-{1\over 3}F_d=<d|Q^{(1)}|d>=-{1\over 2}F_3+{1\over 6}F_8\ ,\eqno(5.3)$$

\noindent with the normalization conditions $F_u(0)=F_d(0)=1$. A combination
of Eq. (5.2) and Eq. (5.3) will result in the quark contributions to the charge 
form factors having the form

$$F_{q,p}={4\over 3}F_u-{1\over 3}F_d\ ,$$
$$F_{q,n}={2\over 3}F_u-{2\over 3}F_d\ .\eqno(5.4)$$

\noindent Since $F_u=1-{r^2_uq^2\over 6}+{\cal O}(q^4)$ and 
$F_d=1-{r^2_dq^2\over 6}+{\cal O}(q^4)$, the quark contributions to 
the proton and neutron mean square charge radii can be calculated as 

$$<r^2>_{q,p}={4\over 3}<r^2>_u-{1\over 3}<r^2>_d\ ,$$ 
$$<r^2>_{q,n}={2\over 3}<r^2>_u-{2\over 3}<r^2>_d\ .\eqno(5.5)$$

If now the $u$- and $d$-quarks are assumed to have the same (mean
square) charge radius Eq. (5.5) can be simplified as

$$<r^2>_{q,p}=<r^2>_u\ ,$$ 
$$<r^2>_{q,n}=0\ .\eqno(5.6)$$

The total mean square charge radius of a nucleon can now be 
calculated as $<r^2>_{tot}=<r^2>_{IA+ex}+<r^2>_{an}+<r^2>_q$, 
where $<r^2>_{IA+ex}$ represents the combined contribution from the impulse
approximation (with relativistic correction) and the necessary exchange 
charge contributions.
The term $<r^2>_{an}$ comes from the anomalous magnetic moment part of
Eq. (2.2) and $<r^2>_q$ is given by Eq. (5.6). 
\par

If the quark contribution $<r^2>_{q,p}$ is chosen so as 
to obtain a value for the total mean square charge radius $<r^2>_{tot,p}$
as close as possible to the empirical one, an estimate of the 
mean square charge radius of the $u$- and $d$-quarks can be done. 
The results in Table II were obtained with the value
$<r^2>_u=<r^2>_d=0.133$ fm$^2$.
\\

\vspace{1cm}

\centerline{\bf VI. DISCUSSION}
\vspace{0.5cm}
 
In quark-quark interaction models that are flavor or velocity dependent
the continuity equation requires the presence of exchange currents. 
The exchange current operators are usually constructed so as to satisfy
the continuity equation to relativistic order $(v/c)^2$. 
The continuity equation to that order does not constrain the corresponding
exchange charge density operators, apart from the requirement that their
contribution vanish with $q$.
In this work the exchange charge density operators that correspond to the 
Fermi-invariant ($SVTAP$) decomposition of the quark-quark interactions
for a two quark system have been constructed (cf. Ref. \cite{BR} where
the corresponding charge density operator for the pseudoscalar $P$ is given
in the case of nucleon-nucleon interactions). Their contribution to the
charge radii of the nucleons has been calculated in a
phenomenological model for the interquark interaction (a version 
of the so called chiral constituent quark model).
The results have then been combined with calculations
in the one-body (impulse) approximation where relativistic corrections 
have been taken into account due to the 
smallness of the (constituent) quark mass to calculate an "intrinsic"
charge radius of the nucleon. This "intrinsic" part can then be added to
the anomalous magnetic moment part of the charge radius to get a total
(mean square) charge radius \cite{EW}. The possibility that 
the constituent quarks are not point-like but have a non-trivial 
electromagnetic structure described by quark form factors has also been 
explored.
\par

A comparison between the results of Table II and the empirical value for 
the electromagnetic charge radius of the proton 
($<r^2>^{1/2}_p=0.862$~fm \cite{Sim1}) shows that the predicted
proton charge radius for the model of Ref. 
\cite{GPVW} without any quark radius contribution is  
9 \% smaller than the empirical value, or 18 \% smaller in the case of the
mean square charge radius. The only exchange 
charge contributions that are considered are the confinement and pseudoscalar
exchange charge contributions.
\par

The result for the neutron electromagnetic mean square charge radius,
on the other hand, has the right sign, and the calculated value of 
$|<r^2>_n|$ is 16 \% larger than the empirical value of $|<r^2>_n|$ 
(empirically $<r^2>_n=-0.117$~fm$^2$ \cite{Dzi}). 
Since there will be no contribution to the neutron charge radius neither
from the impulse approximation nor from confinement (cf. Eq. (4.5), Eq. (4.7)
and Table I), only contributions from pseudoscalar exchange mechanisms 
and the anomalous magnetic moment can affect the sign and the 
numerical value of $<r^2>_n$. 
\par

Without any contributions from the charge radii of the quarks themselves the 
mean square charge radius for the proton  will thus in this model
have a  somewhat smaller value and the neutron a somewhat larger value 
than the empirical ones. If, on the 
other hand, an electromagnetic charge form factor is defined for the
quarks, the expression for the mean square charge radius will contain 
an additional term $<r^2>_q$. If the charge radius for the $u$- and
the $d$-quark is assumed to be equal the proton charge radius will get a
positive contribution, while the neutron charge radius will get no
additional contribution, i.e., $<r^2>_{q,n}=0$.
It is possible to adjust the values for the charge radius of the
quarks to get an agreement between the prediction 
and the empirical data for the proton. When assuming the oscillator
frequency to be 1240~MeV (corresponding to an impulse approximation 
radius without relativistic effects of 0.304~fm \cite{Wagen}), 
the radius for the $u$- and $d$-quarks is determined to be 0.36~fm 
($<r^2>_u=<r^2>_d=0.133$~fm$^2$). 
\par

This can be compared with similar calculations with another flavor
dependent quark-quark interaction model \cite{GPP}, which leads to a
proton mean square charge radius, which is  almost twice as large as 
the empirical one, and to a positive mean square charge radius for the 
neutron if no compensating quark radius term
$<r^2>_q$ is added. To get a reasonable value for the neutron mean
square radius, i.e. not a positive value, without at the same time
causing the proton charge radius to become even larger, would in that model
require the $u$- and $d$-quark radii to differ substantially, $<r^2>_d$ 
being larger than $<r^2>_u$.
\par

The model discussed in this work gives a good description of both the 
spectrum and the magnetic moment of the nucleon. The magnetic 
moments (expressed in nuclear magnetons) for the proton and the neutron,
assuming exchange current contributions only from confinement and 
pseudoscalar exchange mechanisms and using the same  methods for the 
calculations as in Ref. \cite{DGHR}, are 2.58 and -1.77. 
The predicted values for the magnetic moments 
are thus within 8 \% of the experimental results (2.79 and -1.91). 
The parameter describing the confinement strength also seems to be 
realistic in the model. 
The generally accepted value for the string constant of $q\bar q$-systems 
(mesons) is $b\sim 1$~GeV/fm, derived from the charmonium spectrum \cite{Eich} 
and consistent with Regge phenomenology and numerical lattice QCD results 
(for a review article, see Ref. \cite{Lucha}). For baryons the string 
constant is ${1\over 2}b$, which corresponds to the value of the string 
tension parameter $C=2.33$~fm$^{-2}\approx 460$~MeV/fm for the confinement 
in the model studied.\\

\vspace{1cm}
\centerline{\bf ACKNOWLEDGMENT}
\vspace{0.5cm}

The author wishes to thank Prof. D.O. Riska and Dr. L.Ya. Glozman for 
valuable discussions.

\newpage

\centerline{\bf TABLE I.}
\vspace{0.5cm}

\noindent The matrix elements of the spin-flavor operators
for the proton and the neutron:
$Q^{(1)}={1\over 2}[\lambda_3^{(1)}
+{1\over\sqrt{3}}\lambda_8^{(1)}]$, $Q^{(2)}={1\over 2}[\lambda_3^{(2)}
+{1\over\sqrt{3}}\lambda_8^{(2)}]$, 
$Q_\pi^{(12)}={2\over 3}\lambda_3^{(2)}
+\sum_{k=1}^3\lambda_k^{(1)}\lambda_k^{(2)}$,
$Q_\pi^{(21)}={2\over 3}\lambda_3^{(1)}
+{1\over 3}\sum_{k=1}^3\lambda_k^{(2)}\lambda_k^{(1)}$,
$Q_\eta^{(12)}={2\over 3\sqrt{3}}\lambda_8^{(2)}
-{1\over 3}\lambda_8^{(1)}\lambda_8^{(2)}$,
$Q_\eta^{(21)}={2\over 3\sqrt{3}}\lambda_8^{(1)}
-{1\over 3}\lambda_8^{(2)}\lambda_8^{(1)}$,
$Q_{\pi+\eta}^{(12)}={1\over\sqrt{3}}(\lambda_8^{(1)}\lambda_3^{(2)}
+\lambda_3^{(1)}\lambda_8^{(2)})$,
$Q_{\pi+\eta}^{(21)}={1\over\sqrt{3}}(\lambda_8^{(2)}\lambda_3^{(1)}+
\lambda_3^{(2)}\lambda_8^{(1)})$,
$\mbox{\boldmath$\sigma$}^{(1)}\cdot\mbox{\boldmath$\sigma$}^{(2)}
\ Q_\pi^{(12)}$,
$\mbox{\boldmath$\sigma$}^{(2)}\cdot\mbox{\boldmath$\sigma$}^{(1)}
\ Q_\pi^{(21)}$,
$\mbox{\boldmath$\sigma$}^{(1)}\cdot\mbox{\boldmath$\sigma$}^{(2)}
\ Q_\eta^{(12)}$,
$\mbox{\boldmath$\sigma$}^{(2)}\cdot\mbox{\boldmath$\sigma$}^{(1)}
\ Q_\eta^{(21)}$,
$\mbox{\boldmath$\sigma$}^{(1)}\cdot\mbox{\boldmath$\sigma$}^{(2)}
\ Q_{\pi+\eta}^{(12)}$
and $\mbox{\boldmath$\sigma$}^{(2)}\cdot\mbox{\boldmath$\sigma$}^{(1)}
\ Q_{\pi+\eta}^{(21)}$.\\

\begin{center}
\begin{tabular}{|l|c|c|} 
\hline
 && \\
 &\ $p\ $ &\ $n$ \\
 && \\
 \hline
 && \\
$<Q^{(1)}>=<Q^{(2)}>$ & ${1\over 3}$ & 0\\
 && \\
$<\mbox{\boldmath$\sigma$}^{(1)}\cdot\mbox{\boldmath$\sigma$}^{(2)}
\ Q_\pi^{(12)}>
=<\mbox{\boldmath$\sigma$}^{(2)}\cdot\mbox{\boldmath$\sigma$}^{(1)}
\ Q_\pi^{(21)}>$
& ${17\over 9}$ & ${13\over 9}$\\
 && \\
$<\mbox{\boldmath$\sigma$}^{(1)}\cdot\mbox{\boldmath$\sigma$}^{(2)}
\ Q_\eta^{(12)}>
=<\mbox{\boldmath$\sigma$}^{(2)}\cdot\mbox{\boldmath$\sigma$}^{(1)}
\ Q_\eta^{(21)}>$
& $-{1\over 9}$ & $-{1\over 9}$\\
 && \\
$<\mbox{\boldmath$\sigma$}^{(1)}\cdot\mbox{\boldmath$\sigma$}^{(2)}
\ Q_{\pi+\eta}^{(12)}>
=<\mbox{\boldmath$\sigma$}^{(2)}\cdot\mbox{\boldmath$\sigma$}^{(1)}
\ Q_{\pi+\eta}^{(21)}>$
& ${2\over 9}$ & $-{2\over 9}$\\

 && \\
 \hline
\end{tabular}
\end{center}

\newpage

\centerline{\bf TABLE II.}
\vspace{0.5cm}

\noindent Numerical values for the different 
contributions to the mean square charge radius of the proton ($p$) 
and the neutron ($n$) given in fm$^2$  in the model of Ref. \cite{GPVW} when
the impulse approximation radius of the proton (without relativistic 
correction) has the value 0.304~fm ($\omega=1240$~MeV in the harmonic 
oscillator approximation). Here $IA+rel$, $conf$ and
$P$ indicate impulse approximation with relativistic correction, 
confinement and pseudoscalar exchange mechanisms, 
respectively. The contribution $<r^2>_{an}$ comes from the anomalous
magnetic moment part of the electric Sachs form factor, and $<r^2>_{q}$
has been calculated from Eq. (5.6), assuming $<r^2>_u=<r^2>_d=0.133$
fm$^2$.\\

\begin{center}
\begin{tabular}{|l|c|c|} \hline
 && \\
 & $p$ & $n$   \\
 && \\
 \hline
 && \\
 $<r^2>_{IA+rel}$ & 0.345 & 0 \\
 && \\
 $<r^2>_{conf}$ & 0.161 & 0   \\
 && \\
 $<r^2>_P$ & $-$0.015 & $-$0.009   \\
 && \\
 $<r^2>_{an}$ & 0.119 & $-$0.127 \\
 && \\
 $<r^2>_q$ & 0.133 & 0  \\
 && \\
 \hline
 && \\
 $<r^2>_{tot}$ & 0.743 & $-$0.136\\
 && \\
 $<r^2>_{exp}$ & (0.862)$^2$ & $-$0.117  \\
 && \\
 \hline
\end{tabular}
\end{center}
\end{document}